\documentclass[reprint,amsmath,amssymb,aps,prd]{revtex4-1}

\usepackage{graphicx}
\usepackage{aas_macros}
\usepackage[usenames, dvipsnames]{color}

\newcommand{\dd}{\mathrm{d}}
\newcommand{\chieff}{\chi_\mathrm{eff}}
\newcommand{\isoOdds}{\mathcal{O}_\mathrm{iso}}
\newcommand{\aliOdds}{\mathcal{O}_\mathrm{ali}}

\newcommand{\aliOddsVal}{1.1}

\begin{document}

\title{Using spin to understand the formation of LIGO's black holes}

\author{Ben Farr}
 \email{bfarr@uoregon.edu}
\affiliation{Department of Physics, University of Oregon, Eugene, OR 97403, USA \\
Enrico Fermi Institute, University of Chicago, Chicago, IL 60637, USA \\
Kavli Institute for Cosmological Physics, University of Chicago, Chicago, IL 60637, USA \\ and
DARK, Niels Bohr Institute, University of Copenhagen, Blegdamsvej 17, 2100 Copenhagen, Denmark}
\author{Daniel E. Holz}
 \affiliation{Kavli Institute for Cosmological Physics, University of Chicago, Chicago, IL 60637, USA \\
Department of Physics, University of Chicago, Chicago, IL 60637, USA \\
Enrico Fermi Institute, University of Chicago, Chicago, IL 60637, USA \\
Department of Astronomy and Astrophysics, University of Chicago, Chicago, IL 60637, USA \\ and
DARK, Niels Bohr Institute, University of Copenhagen, Blegdamsvej 17, 2100 Copenhagen, Denmark}
\author{Will M. Farr}
\affiliation{Birmingham Institute for Gravitational Wave Astronomy and School of Physics and Astronomy, University of Birmingham, Birmingham, B15 2TT, United Kingdom \\ and
DARK, Niels Bohr Institute, University of Copenhagen, Blegdamsvej 17, 2100 Copenhagen, Denmark}

\date{\today}

\begin{abstract}
    With the detection of four candidate binary black hole (BBH) mergers by the Advanced LIGO detectors thus far, it is becoming possible to constrain the properties of the BBH merger population in order to better understand the formation of these systems.  Black hole (BH) spin orientations are one of the cleanest discriminators of formation history, with BHs in dynamically formed binaries in dense stellar environments expected to have spins distributed isotropically, in contrast to isolated populations where stellar evolution is expected to induce BH spins preferentially aligned with the orbital angular momentum.  In this work we propose a simple, model-agnostic approach to characterizing the spin properties of LIGO's BBH population.  Using measurements of the effective spin of the binaries, which is LIGO's best constrained spin parameter, we introduce a simple parameter to quantify the fraction of the population that is isotropically distributed, regardless of the spin magnitude distribution of the population. Once the orientation characteristics of the population have been determined, we show how measurements of effective spin can be used to directly constrain the underlying BH spin magnitude distribution.
Although we find that the majority of the current effective spin measurements are too small to be informative, with LIGO's four BBH candidates we find a slight preference for an underlying population with aligned spins over one with isotropic spins (with an odds ratio of \aliOddsVal).
We argue that it will be possible to distinguish symmetric and anti-symmetric populations at high confidence with tens of additional detections, although mixed populations may take significantly more detections to disentangle.
We also derive preliminary spin magnitude distributions for LIGO's black holes, under the assumption of aligned or isotropic populations.
\end{abstract}

\maketitle

\section{\label{sec:intro}Introduction}
The direct detection of gravitational waves (GWs) from binary black holes (BBHs) has become almost routine, with three confident events (GW150914 \citep{gw150914}, GW151226 \citep{gw151226}, and GW170104 \citep{gw170104}) and one candidate event LVT151012 \citep{o1bbh} identified by the Advanced LIGO detectors.  The GW signatures of these binaries encode properties of the binary \citep{lalinference}, and in particular can be used to measure the spin properties of the binary's black holes (BHs).  The astrophysical processes by which these systems form remain uncertain; two generic formation channels include the evolution of an isolated pair of stars that were born together (i.e., ``in the field'') \citep{1993MNRAS.260..675T,2016Natur.534..512B,2017NatCo...814906S,2016MNRAS.458.2634M,2016A&A...588A..50M}, and dynamical interactions in a dense stellar environment (i.e., globular clusters) \citep{1993Natur.364..423S,2000ApJ...528L..17P,2015PhRvL.115e1101R,2016ApJ...832L...2R}.

Computational models of these formation channels provide predictions for the rates and distributions of binary masses. However, these predictions are highly dependent upon assumptions about poorly understood processes (e.g., common envelope evolution).  More robust are the predictions for binary spin properties, particularly the {\em orientation}\/ of BH spins with respect to the orbital angular momentum of the binary. A generic characteristic of dynamical formation is that the spin orientations of the component black holes are isotropic with respect to the orbital angular momentum \cite{SigurdssonHernquist:1993,PZMcMillan:2000,Rodriguez:2015,Stone:2016,2016ApJ...832L...2R,2004PhRvD..70l4020S,2007ApJ...661L.147B}. Isolated binaries, on the other hand, are generally expected to be preferentially aligned with the orbital angular momentum \cite{TutukovYungelson:1993,2016Natur.534..512B,Stevenson:2017,MandeldeMink:2016,Marchant:2016}.This conclusion can be weakened through the impact of natal kicks \cite{ROSkicks}, but it is difficult for the isolated channel to produce a significant fraction binary mergers with large misalignment ($\hat{S} \cdot \hat{L} < 0$, where $\hat{S}$ and $\hat{L}$ point along the stellar spin and orbital angular momentum vectors, respectively).
In the absence of firm predictions for the full spin distributions (orientation and magnitude) from various formation channels, we provide a model-agnostic approach specially suited for looking for spin isotropy in the BBH population.

LIGO's spin measurements come with large, but well-quantified, uncertainties.  Thus it is difficult to utilize spin to constrain the formation channel of any particular binary, but with many events the properties of the population can start to be inferred.  In principle there are two astrophysically interesting spin quantities for each of the component black holes: the spin magnitude and its misalignment angle with the orbital angular momentum.  However, LIGO typically provides only one well-constrained spin quantity: the \emph{effective spin}, $\chieff$, which is the mass-weighted combination of the aligned components of BH spins \cite{gw150914_pe}:
\begin{equation}
\label{eq:chieff-def}
\chieff \equiv \frac{1}{M} \left( m_1 \chi_1 + m_2 \chi_2 \right),
\end{equation}
where $M = m_1 + m_2$ is the total mass of the system, and $-1 < \chi_{1,2} < 1$ are the projections of the spin vectors  of each component BH along the orbital axis.  As pointed out in \citep{farrnature}, by measuring the effective spin distribution of the population we can infer misalignment characteristics, and consequently formation channels.

Where \citet{farrnature} focused on assessing the plausibility of specific model populations through $\chieff$ measurements, we generalize this approach to make model-agnostic statements about the misalignment characteristics of the population.  Our approach is based on the basic principle that \emph{any isotropic population, regardless of the population's spin magnitude distribution, must have a symmetric $\chieff$ distribution}.  Thus we can use a hierarchical analysis of LIGO's detected events to compare the relative rates of binaries with $\chieff < 0$ and $\chieff > 0$ to assess whether or not the detected population has spins that are isotropically distributed.  If $\rho$ is the fraction of systems with $\chieff > 0$, then the isotropic (or symmetric) fraction is $\sim 2(1-\rho)$ and the aligned fraction is $\sim 2\rho - 1$.  A single definitive measurement of $\chi_{\rm eff}<0$ would imply that $\rho<1$; to date no such measurement exists. Section \ref{sec:methods} presents our hierarchical analysis framework, our proposed models for the $\chieff$ distribution, and a prescription for determining the underlying BH spin magnitude distribution once the orientation distribution has been characterized.  Section \ref{sec:results} presents the results from applying these techniques to LIGO's current catalog of BBH candidates.

\section{\label{sec:methods}Methods}
\subsection{\label{sec:simulations}Simulated Populations}
To test the sensitivity of the techniques outlined in this paper we consider several fiducial distributions. Our approach is similar to \citet{farrnature}.  We will simulate isotropic and aligned populations with four different component spin magnitude distributions: a ``high-spin'' distribution with $p(a) = 2a$, a ``low-spin'' distribution with $p(a) = 2(1-a)$, a ``very-low-spin'' population with $a\sim |\mathcal{N}(0, 0.1)|$, and a ``very-very-low-spin'' population with $a\sim |\mathcal{N}(0, 0.01)|$, where $\mathcal{N(\mbox{mean},\mbox{standard deviation})}$ represents a normal distribution. For the purposes of simulating aligned populations we assume exact alignment of component spins and the orbital angular momentum, but the techniques described here are general, and can be applied to populations with preferential (but not exact) alignment.  In particular, unless the processes that misalign an isolated (field) population can produce anti-aligned spin vectors ($\hat{L} \cdot \hat{S} < 0$) the population will still have $\chieff > 0$.  Posterior constraints on $\chieff$ for each event are simulated to be Gaussian with widths assigned randomly from LIGO's current BBH constraints \citep{o1bbh,gw170104}.

\subsection{\label{sec:h-bayes} Hierarchical Population Inference}
Ultimately we would like to calculate $p(\lambda\rvert\{d^i\})$, the posterior density function for the parameters $\lambda$ describing our population given the data measured around each event $\{d^i\}$.  Since the effective spin cannot be determined precisely form any given observation, we want to marginalize over it, leading to the following expression for the posterior density function for population parameters $\lambda$:
\begin{equation}
\label{eqn:h-bayes}
p\left(\lambda \rvert\left\{d^i\right\}\right) \propto \left[\prod_{i=1}^{N_\mathrm{obs}}\int \dd\chieff^i p\left(d^i\rvert\chieff^i\right)p\left(\chieff^i\rvert\lambda\right)\right]p(\lambda),
\end{equation}
where $p(d^i|\chieff^i)$ is the likelihood of measuring data $d^i$ given $\chieff^i$, $p(\chieff^i|\lambda)$ is the probability of observing $\chieff^i$ given population parameters $\lambda$, and $p(\lambda)$ is the prior probability of observing population parameters $\lambda$.

Given $N_i$ samples $\{\chieff^{ij}\}$ from the marginal posteriors for event $i$ generated using a prior $p\left( \chieff \right)$ we can avoid reanalysis of the data by reweighing the individual event posteriors given a particular population model, leading to a more practical form of the posterior for model parameters
\begin{equation}
\label{eqn:hpost}
p\left(\lambda \rvert\left\{d^i\right\}\right) \propto \left[\prod_{i=1}^{N_\mathrm{obs}} \frac{1}{N_i}\sum_{j=1}^{N_i}\frac{p\left(\chieff^{ij}\rvert\lambda\right)}{p\left(\chieff^{ij}\right)} \right] p(\lambda).
\end{equation}
We will make use of this hierarchical posterior density function throughout this work using various population models.

\subsection{\label{sec:two-bin}Two-bin $\chieff$ model}
Since we do not have concrete predictions for the expected distribution of spin parameters for various formation channels, we want a simple yet general approach to characterize the binary population that does not depend on unknown properties, such as the spin magnitude distribution.  Binaries that form dynamically, regardless of their spin magnitude distribution, are expected to have a $\chieff$ distribution that is symmetric about 0.  A simple way to measure this is to use a two-bin, one parameter model to describe the $\chieff$ distribution,
\begin{equation}
\label{eqn:two-bin}
p(\chieff) = \begin{cases}
1 - \rho ~~ -1\leq \chieff < 0 &\\
\rho \hfill 0 \leq \chieff \leq 1 &
\end{cases},
\end{equation}
where $\rho$ can be interpreted as the fraction of systems with $\chieff>0$.  Under this model purely isotropic populations have $\rho=1/2$, aligned populations have $\rho = 1$, and purely anti-aligned populations ($\chieff<0$) have $\rho = 0$. In this work we simulate populations with perfect alignment ($\rho=1$), and while natal kicks are expected to introduce some misalignment in the isolated population, systems with $\chieff<0$ are expected to be rare since this requires torquing the black hole all the way over to retrograde spin. Thus $\rho$ is expected to be \emph{very}\/ close to $1$ for isolated populations \citep{ROSkicks}.

If one believes that the BBHs detected by LIGO have to either come from a perfectly isotropic population \emph{or} a perfectly aligned population, then the odds ratio in favor of isotropy (assuming equal prior odds) goes as $\isoOdds = p(\rho=1/2)/p(\rho=1)$.  If we consider the possibility of a mixed population of binaries, then the posterior density function for $\rho$ provides a direct measure of the mixing fraction, with $f_\mathrm{aligned} = 2\rho - 1$ \cite{Vitale2017,Stevenson2017Spins,TalbotThrane2017spins}.

\begin{figure}[ht!]
  \centering
  \includegraphics[width=0.45\textwidth]{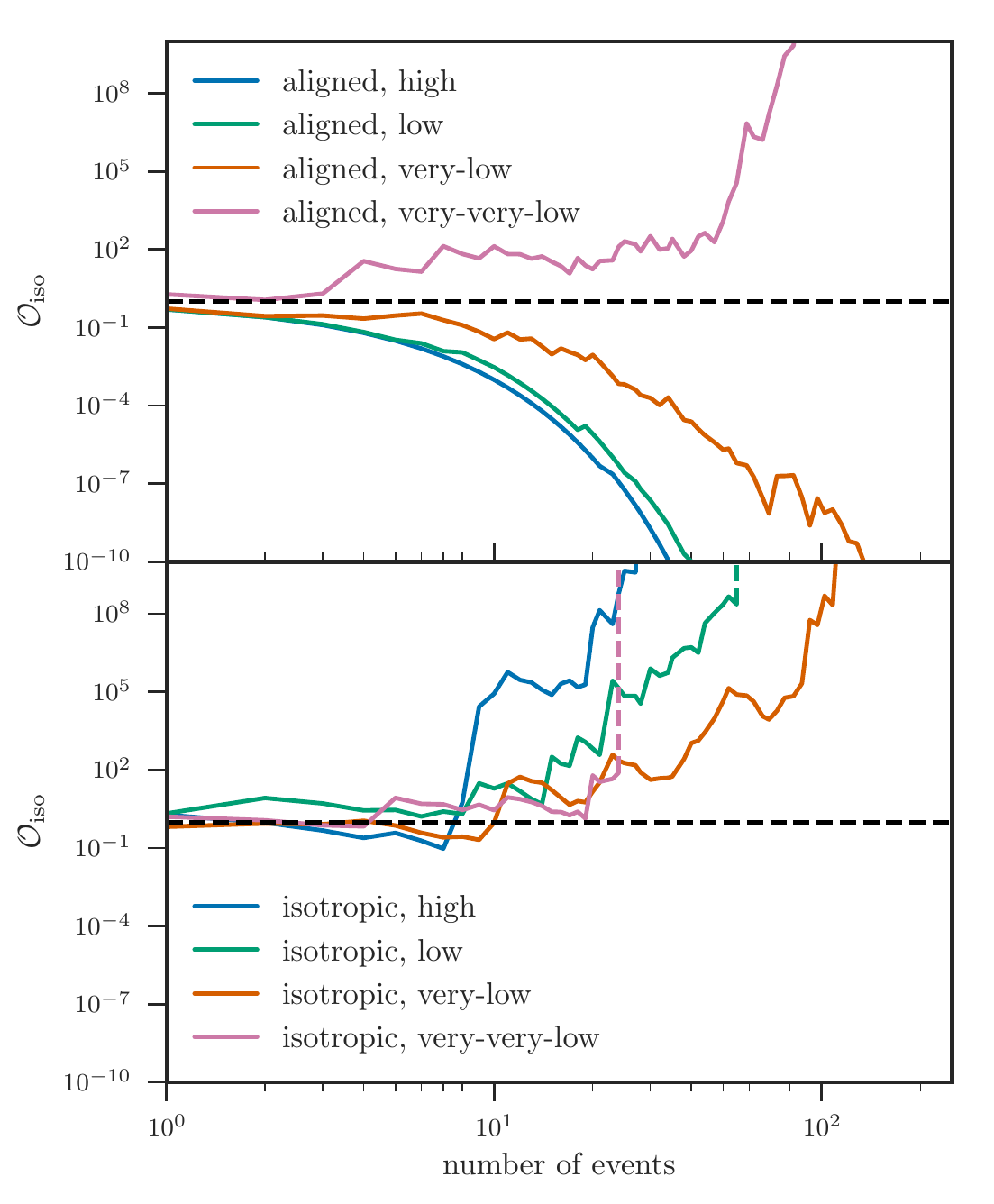}
    \caption{\label{fig:odds_ratio}Evolution of the odds ratio in favor of an isotropic distribution (where $\mathcal{O}_\mathrm{iso} > 1$ indicates an isotropic distribution is preferred over aligned) for simulated populations using the two-bin model, applied to aligned populations (\emph{top}) and isotropic populations (\emph{bottom}).  Dashed lines indicate an infinite odds ratio, occurring when a definitive $\chieff<0$ measurement is made. If the population's spin magnitudes are \emph{very} low ($a\lesssim0.01$), the two-bin model breaks down and can wrongly conclude that a population is isotropically distributed. This limitation is addressed with the 3-bin model introduced in the next section.
  }
\end{figure}
To test the effectiveness of the two-bin model we simulate up to $250$ events from isotropic and aligned spin populations with observational uncertainties drawn randomly from LIGO's current BBH catalog for the various spin magnitude distributions outlined in section \ref{sec:simulations}.  To summarize the performance of this model, Figure~\ref{fig:odds_ratio} shows how the odds ratio $\isoOdds$ in favor of an isotropic population estimated using the two-bin model evolves as an increasing number of events are detected from each population.  For isotropic populations the odds ratio behaves as expected, increasing steadily with the number of detections.  For aligned populations with very-very-low-spin, however, the odds ratio in fact favors the wrong conclusion (i.e., the pink curve in the top panel of Fig.~\ref{fig:odds_ratio} goes up, not down).
\begin{figure}[ht!]
  \centering
  \includegraphics[width=0.5\textwidth]{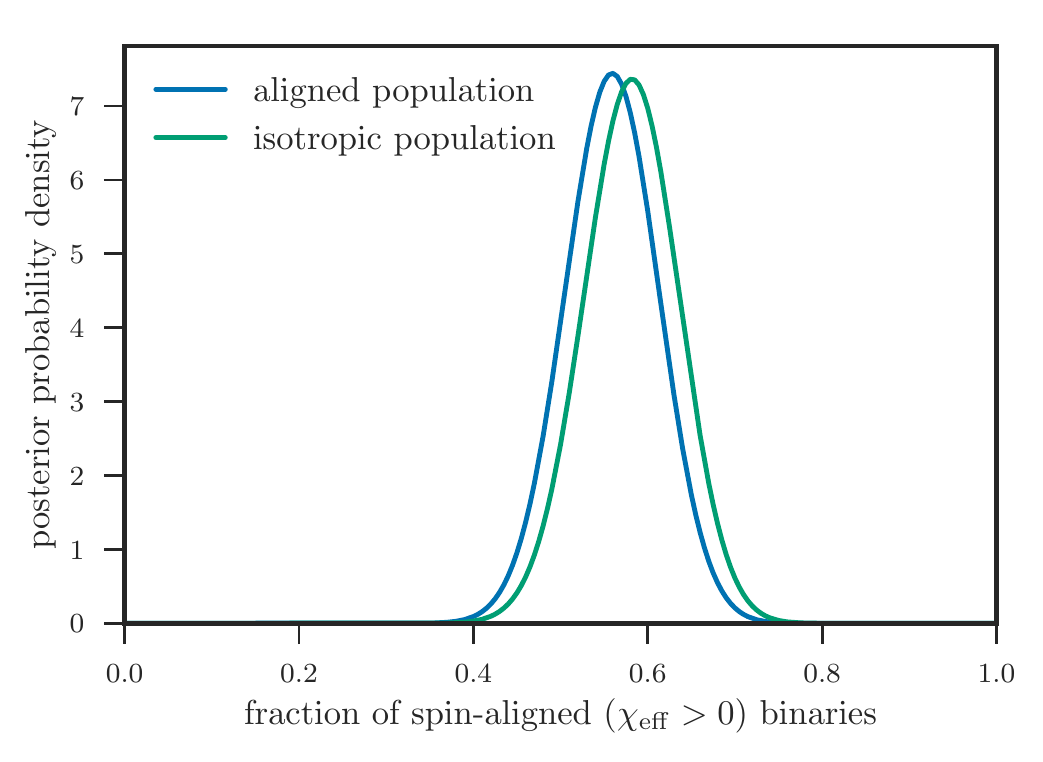}
    \caption{\label{fig:two-bin-vvlow}Posterior distributions for the population parameter $\rho$, the fraction of systems with $\chieff>0$, after $250$ simulated events from the isotropic and aligned, very-very-low-spin ($a\sim |\mathcal{N}(0, 0.01)|$) populations.  If the detected population has such low spins, the two-bin model fails (see section ~\ref{sec:two-bin}), motivating the use of the three-bin model (see section ~\ref{sec:three-bin}).}
\end{figure}
Figure~\ref{fig:two-bin-vvlow} shows the posterior constraints on $\rho$ for the very-very-low-spin aligned and isotropic populations, where for the aligned population the simulated value ($\rho=1$) is incorrectly excluded with high confidence. This shortcoming, due to the mismatch between the population model and the simulated population, is addressed in the following subsection.

For populations with small spins, we have $w \ll \sigma_\mathrm{obs} \ll 1$, where $w$ is the characteristic width of the population about $\chieff = 0$ and $\sigma_\mathrm{obs}$ is the typical observational uncertainty of $\chieff$.  In this case the log posterior for $\rho$ becomes (c.f.\ Eq. \eqref{eqn:h-bayes})
\begin{multline}
\log p\left( \rho \mid \left\{ d^i \right\} \right) = \mathrm{const} \\ + \sum_i 2\left[ \left( A_i - \frac{1}{2} \right) \left( 2 \rho - 1\right) - \left(A_i - \frac{1}{2} \right)^2 \left( 2\rho - 1 \right)^2 \right. \\ \left. + \mathcal{O}\left( A_i - \frac{1}{2}\right)^3 \right]
\end{multline}
where
\begin{equation}
A_i = \frac{\int_0^1 \dd \chi \, p\left( d^i \mid \chi \right)}{\int_{-1}^1 \dd \chi \, p\left( d^i \mid \chi \right)}
\end{equation}
is the fraction of the likelihood that supports $\chieff > 0$.  Thus
\begin{multline}
\label{eq:approx-posterior-near-zero}
\log p\left( \rho \mid \left\{ d^i \right\} \right) \simeq \mathrm{const} \\ + 2 N \left( \mu \left(2\rho - 1\right) - \sigma^2 \left(2 \rho -1\right)^2 \right),
\end{multline}
where
\begin{equation}
\mu \equiv \left\langle A - \frac{1}{2} \right\rangle
\end{equation}
and
\begin{equation}
\sigma^2 \equiv \left\langle \left( A - \frac{1}{2} \right)^2 \right\rangle.
\end{equation}
The posterior in Eq. \eqref{eq:approx-posterior-near-zero} peaks at
\begin{equation}
\hat{\rho} = \frac{\mu + 2 \sigma^2}{4 \sigma^2}.
\end{equation}
For the aligned, very-very-low-spin population considered in this work, $\mu \simeq 0.022$ and $\sigma \simeq 0.29$, leading to $\hat{\rho} \simeq 0.57$, as observed in Figure \ref{fig:two-bin-vvlow}, and explaining why the correct (aligned) model is not preferred even as the number of measurements increases in Figure \ref{fig:odds_ratio}.  Any population where $\chieff$ is typically much smaller than the observational uncertainty will be subject to the same effect in the two-bin model, motivating our introduction of a three-bin model in the next subsection.

\subsection{\label{sec:three-bin}Three-bin $\chieff$ model}
The two-bin approach outlined in Subsection~\ref{sec:two-bin} is a simple approach to answering the question of isotropy.  However, systems with $\chieff \approx 0$ will fall on the boundary between the two bins, and are therefore uninformative about $\rho$ (the fraction of the population with $\chieff > 0$).  When the typical width of the population is smaller than the observational uncertainty the two-bin model breaks down, as explained in Section \ref{sec:two-bin}.  To be more robust to the case of low-$\chieff$ systems, we introduce a third bin to the model described in section~\ref{sec:two-bin}. This additional bin is narrow and centered on $\chieff=0$, leaving us with a two-parameter model:
\begin{equation}
\label{eqn:three-bin}
p(\chieff) = \begin{cases}
2\alpha  (1 - \rho)/(2 - \Delta) & -1 \leq \chieff < -\frac{\Delta}{2} \\
(1 - \alpha)/\Delta & -\frac{\Delta}{2} \leq \chieff < \frac{\Delta}{2} \\
2\alpha \rho/(2 - \Delta) & -1 \leq \chieff \leq -\frac{\Delta}{2} \\
\end{cases},
\end{equation}
where $\Delta$ is the width of the central bin, $\alpha$ is the fraction of ``informative'' systems that lie outside of the central (low-$\chieff$) bin, and $\rho$ is the fraction of those systems with $\chieff>0$.  This third, narrow bin accounts for systems with $\chieff$ too small to be informative.  By marginalizing over the number of informative systems, we obtain a marginal posterior distribution function $p(\rho)$ that can be interpreted in the same way as the two-bin model.  We have experimented with several choices for the central bin width $\Delta$.  Since our results do not depend sensitively on this choice, we have used $\Delta = 0.05$ throughout this work; this width is smaller than, but comparable to, the typical observational uncertainty for existing LIGO events.

\begin{figure}[t!]
  \centering
  \includegraphics[width=0.5\textwidth]{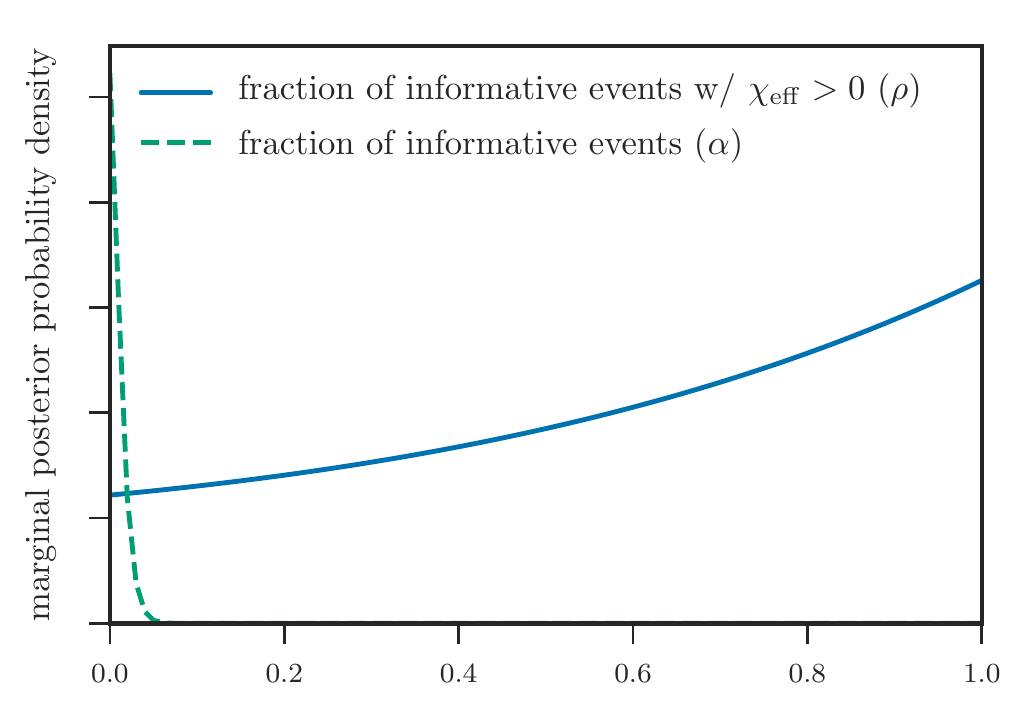}
    \caption{\label{fig:three-bin-vvlow}Marginal posterior distributions for the fraction of informative systems $\alpha$ and fraction of informative systems with $\chieff>0$, $\rho$, after $250$ events for the aligned, very-very-low-spin population.  In contrast to Figure \ref{fig:two-bin-vvlow}, almost all systems are uninformative, and leaving $\rho$ only marginally constrained.}
\end{figure}
\begin{figure}[t!]
  \centering
  \includegraphics[width=0.5\textwidth]{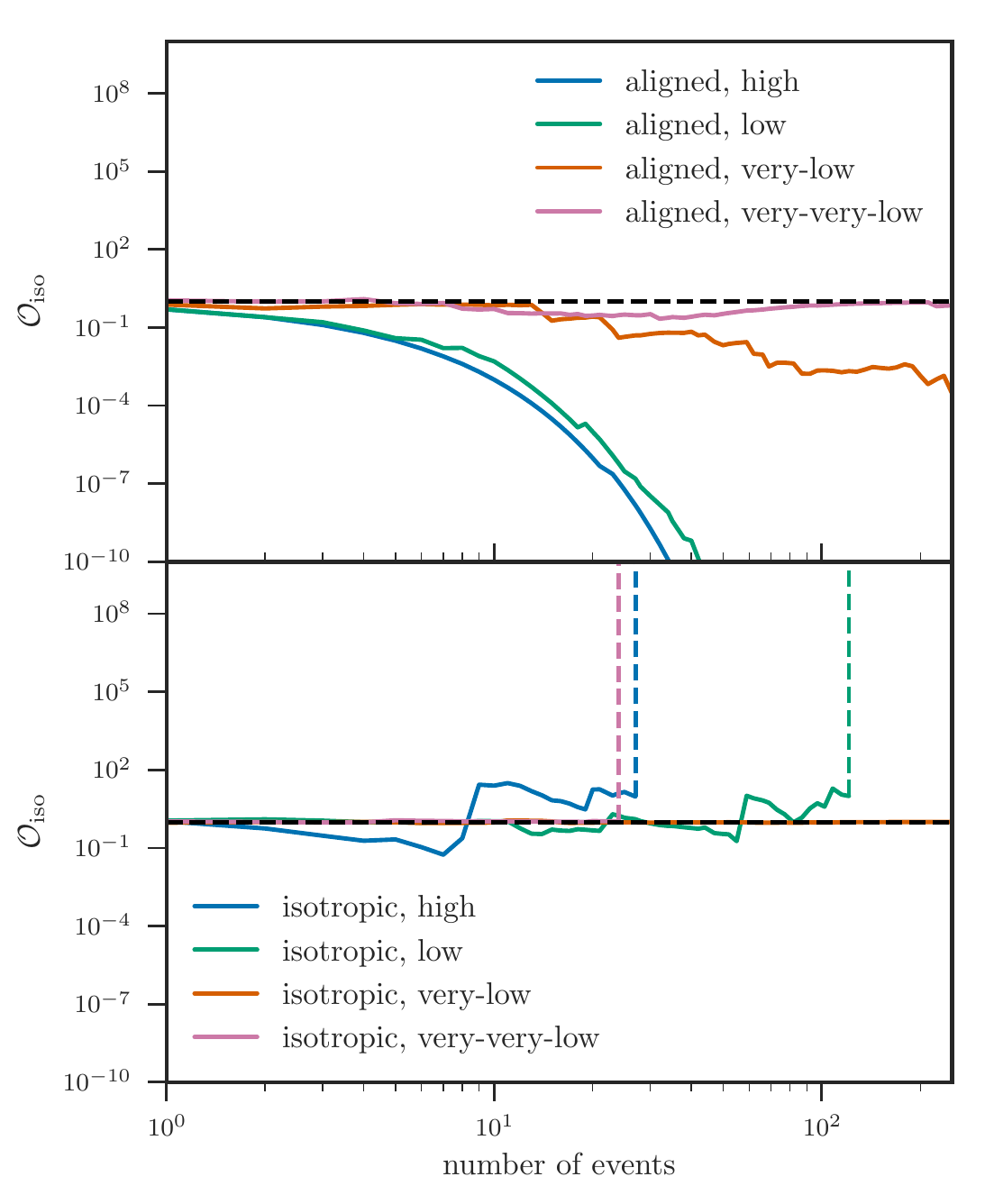}
  \caption{\label{fig:3bin_odds_ratio}Evolution of the odds ratio in favor of an isotropic population over an aligned population for the various simulated populations using the three-bin model, applied to aligned populations (\emph{top}) and isotropic populations (\emph{bottom}).  Dashed lines indicate an infinite odds ratio, occurring when a definitive $\chieff<0$ measurement is made. Unlike the 2-bin model presented in Fig.~\ref{fig:odds_ratio}, this model is robust against very-very-low-spin populations.}
\end{figure}
Figure~\ref{fig:three-bin-vvlow} shows the results from applying the three-bin model to the same very-very-low-spin aligned population that failed with the two-bin model (see Fig. \ref{fig:odds_ratio}).  The marginal posterior distribution for $\alpha$ (the fraction of systems lying outside of the low-$\chieff$ bin) shows a vanishing number of events to be informative, while the marginal posterior distribution for $\rho$ (the fraction of informative systems with $\chieff > 0$) remains unconstrained.  Figure~\ref{fig:3bin_odds_ratio} shows how the odds ratio in favor of an isotropic distribution, calculated from the marginal posterior distribution of $\rho$, performs on isotropic and aligned populations. We find that tens of events can characterize the underlying population at high confidence, with the precise number depending on the fraction of the population with effective spins indistinguishable from zero.

\subsection{\label{sec:spin-mag}Constraining Component Spins}
\begin{figure}[bt!]
  \centering
  \includegraphics[width=0.5\textwidth]{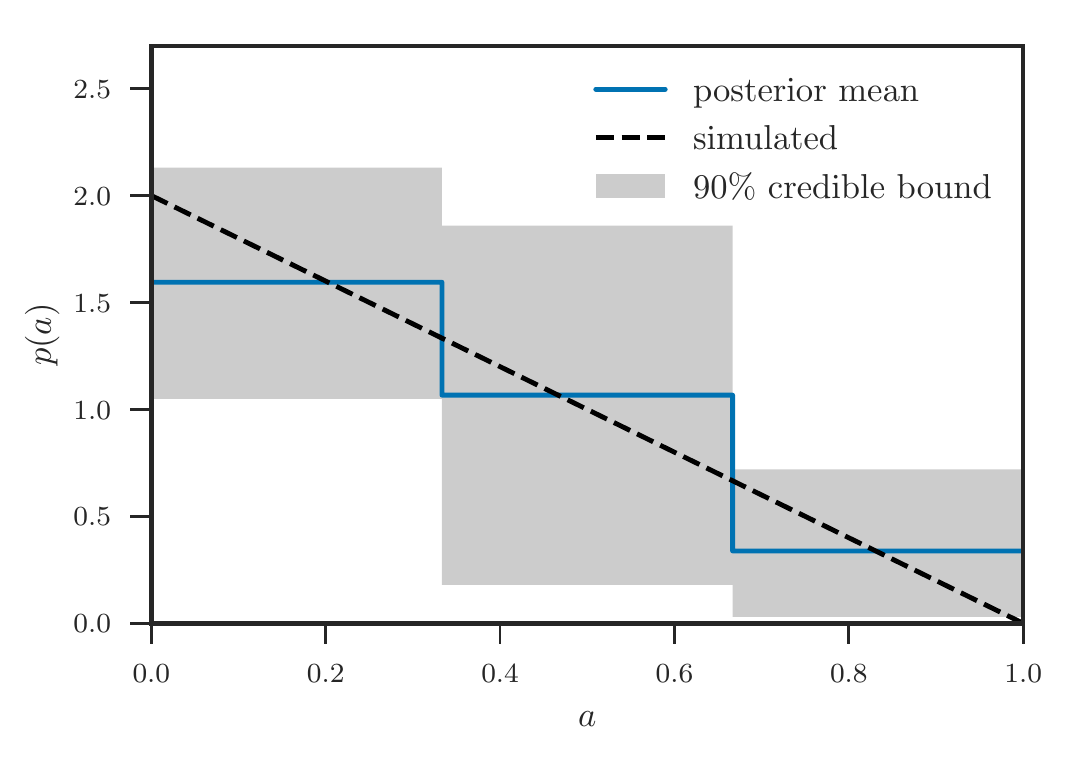}
\caption{\label{fig:iso-low-spin-mag}Mean and $90\%$ credible bounds on the spin magnitude distribution of the low-spin isotropic population after 250 detections, under the assumption that the population is isotropic.}
\end{figure}

In general it is difficult to infer the spin magnitude distribution for the component black holes from the measured distribution of $\chieff$, since the relation between the two distributions depends on mass ratios and spin alignments. However, using our simple method to identify the alignments of the population can dramatically reduce the complexity of this task.
If the population has been determined to be either completely isotropic or aligned, we can then use $\chieff$ measurements to constrain the underlying BH spin magnitude distribution.  We will use a three-bin model for the spin magnitude distribution
\begin{equation}
\label{eqn:mag_dist}
p(a) = \begin{cases}
A_1/3 & 0 \leq a < 1/3 \\
A_2/3 & 1/3 \leq a < 2/3 \\
(1 - (A_1 + A_2))/3 & 2/3 \leq a \leq 1
\end{cases},
\end{equation}
where $A_1$ and $A_2$ are the heights of the low ($0 \leq a < 1/3$) and moderate ($1/3 \leq a < 2/3$) spin bins, respectively.  With this model in hand, assuming either an isotropic or aligned distribution of orientations leads to a unique prediction for the distribution of $\chieff$ that can be used in place of the two and three bin models described above.
As an example, Figure~\ref{fig:iso-low-spin-mag} shows the posterior constraints on the spin magnitude distribution after detecting 250 events from a low-spin isotropic population.

\section{\label{sec:results}Results from LIGO's BBHs}
We now make inferences about the spin distributions of the BBH population being detected by LIGO.
In lieu of true posterior samples from the LIGO analyses, which are not publicly available, we approximate the posterior estimates of $\chieff$ for the four likely GW events detected so far (GW150914, GW151226, GW170104, and LVT151012) following a similar prescription to that in \citep{farrnature}.  We approximate the posterior as a Gaussian whose central 90\% credible interval matches the stated interval for each event \citep{o1bbh,gw170104}.
\begin{figure}[ht!]
  \centering
  \includegraphics[width=0.5\textwidth]{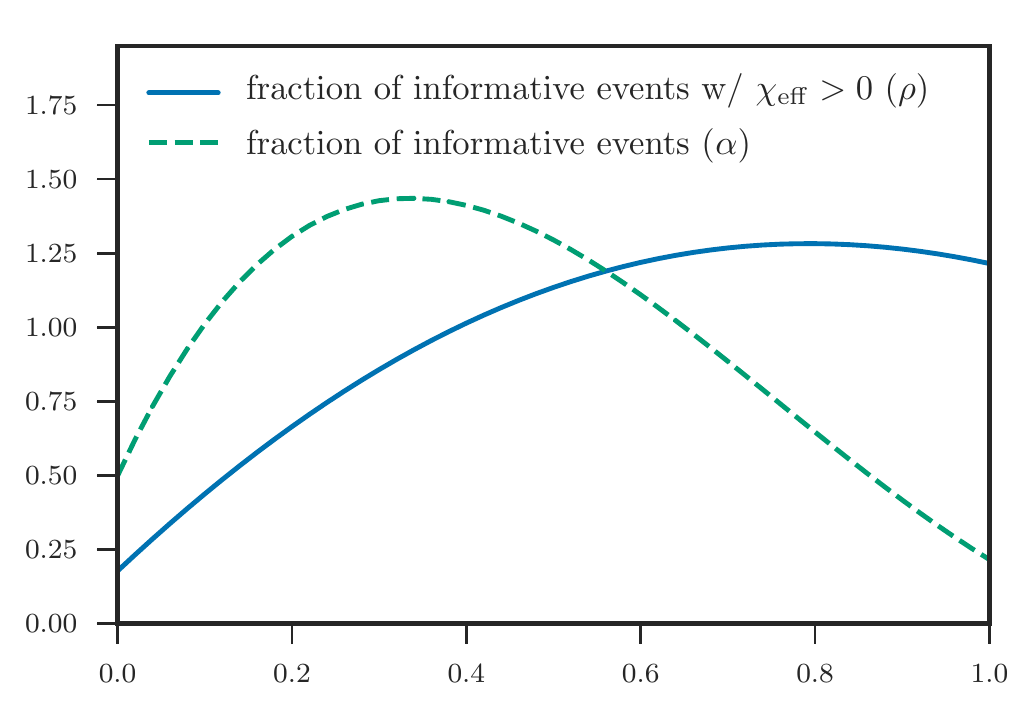}
    \caption{\label{fig:mock_real_marginal}Marginal posterior distributions for the fraction of informative systems $\alpha$ and fraction of informative systems with $\chieff>0$, $\rho$, for the $4$ likely GW events detected by LIGO so far, indicating an odds ratio of $\aliOdds=\aliOddsVal$, i.e., marginal support for an aligned population.}
\end{figure}

Figure~\ref{fig:mock_real_marginal} shows the marginal posterior density functions from a three-bin analysis of LIGO's four likely BBH detections.
We find that the information for distinguishing between symmetric and anti-symmetric spin distributions is dominated by GW151226; the rest of the events are consistent with the low-$\chieff$ central bin. Roughly speaking, in the symmetric case a nonzero $\chieff$ measurement would have a 50-50\% change of being aligned ($\chieff>0$) or anti-aligned ($\chieff<0$). In the aligned case, all of these systems would have $\chieff>0$. Thus finding GW151226 with $\chieff>0$, and no definitive systems with $\chieff<0$, weakly favors aligned versus anti-aligned, if those are the only two possibilities. Assuming that the current results are ``representative'', roughly speaking we expect every sample of $\sim4$ events to contain one with $|\chieff|\ne0$ at high confidence. It will rapidly become evident whether these are evenly distributed between positive and negative $\chieff$ values, or whether they are preferentially positive. A single confident measurement of $\chieff<0$ will invalidate the asymmetric population assumption (i.e., will imply $\rho\ne1$).

\begin{figure}[t!]
  \centering
  \includegraphics[width=0.5\textwidth]{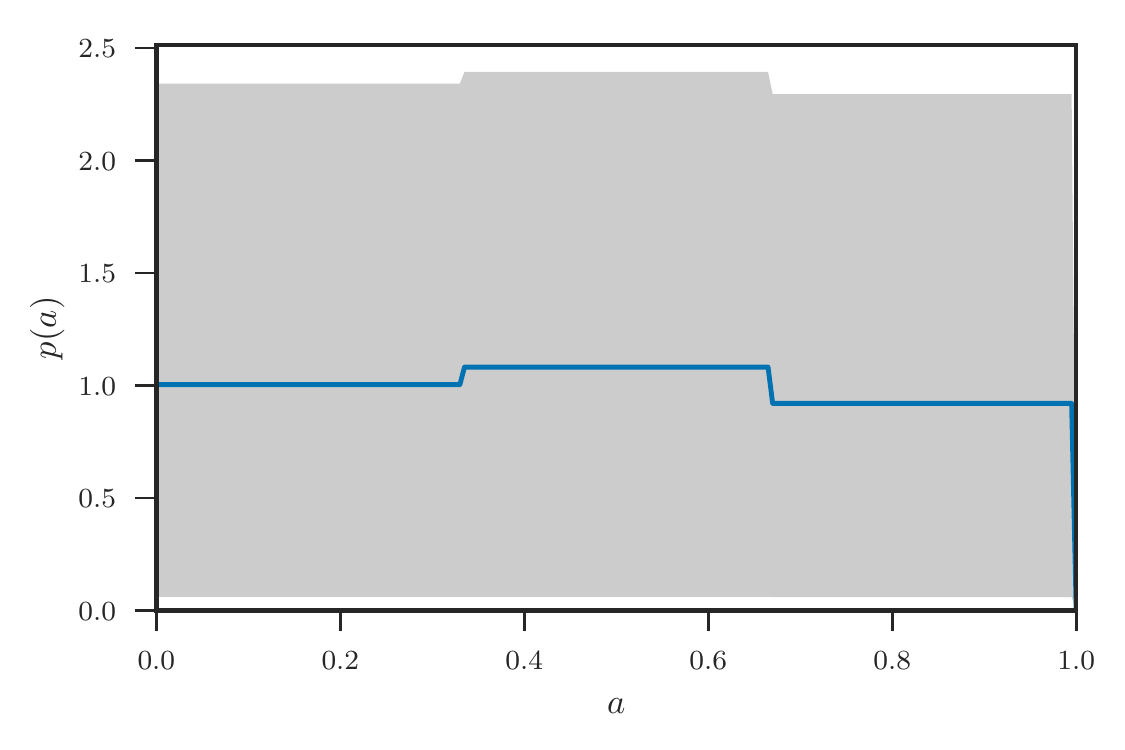}
  \caption{\label{fig:mock-iso-mag}Mean and $90\%$ credible bounds on the spin magnitude distribution inferred from LIGO's $4$ likely BBH detections, assuming the population is isotropically distributed.}
\end{figure}

\begin{figure}[tb!]
  \centering
  \includegraphics[width=0.5\textwidth]{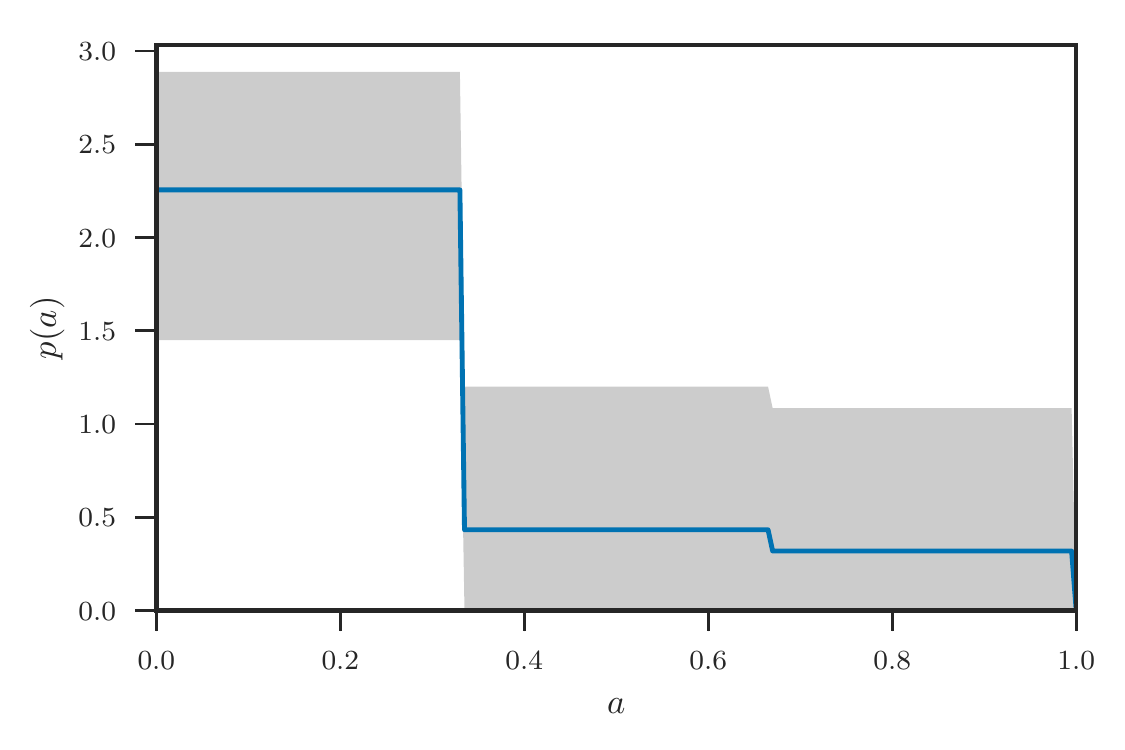}
  \caption{\label{fig:mock-ali-mag}Mean and $90\%$ credible bounds on the spin magnitude distribution inferred from LIGO's $4$ likely BBH detections, assuming the population is aligned.}
\end{figure}

If we now assume a particular formation scenario for LIGO's BBHs, we can infer the BH spin magnitude distribution following the approach in Section~\ref{sec:spin-mag}.  Figure~\ref{fig:mock-iso-mag} shows the posterior constraints on the spin magnitude distribution assuming all of LIGO's BBHs were dynamically formed.  Figure~\ref{fig:mock-ali-mag} shows the posterior constraints on the spin magnitude distribution assuming all of LIGO's BBHs are aligned, the currently preferred scenario.  If the population is indeed aligned, then even with just the 4 candidate BBH mergers detected to date we can already say that there are likely more low spin systems ($a\lesssim1/3$) than moderate ($1/3\lesssim a\lesssim2/3$) or high ($a\gtrsim2/3$) spin systems.

A maximal spin ($a \equiv 1$) aligned model is strongly disfavored, with an odds ratio compared to our three-bin model of $\sim 10^{-39}$, while under an isotropic assumption the maximal spin model is disfavored by a modest odds ratio of $0.33$; the striking difference between the aligned and isotropic assumptions reflects the difficulty in distinguishing different spin magnitude models under the isotropic assumption \cite{farrnature}.  Our approach considers only $\chieff$, and neglects information about the precessing component of the spins, $\chi_p$. In the case of maximal spins, however, the typically large precessing components of the spins are likely to be measurable, and thus the odds ratio against a maximal, isotropic spin population from the full GW data set is probably larger than the 3:1 value quoted above.

\section{Conclusions}
We have described a simple, model-agnostic approach for distinguishing dynamically formed and isolated populations from spin measurements ($\chieff$) of LIGO's events.
Our model asks whether the $\chieff$ distribution is symmetric (suggesting a dynamical formation channel), or whether there is evidence for asymmetry (i.e., a preference for aligned versus anti-aligned systems, suggesting isolated formation). We introduce a mixing fraction, $\rho$, where $\rho=0.5$ is isotropic, $\rho=1$ is 100\% aligned, and a value in between represents a mixed population. Furthermore, we show that once the alignment of the population has been characterized, the spin magnitude distribution can be directly inferred from the $\chieff$ distribution.

With the 4 likely BBH systems observed by LIGO thus far we find that an aligned formation scenario (i.e., isolated or field formation) is slightly preferred over an isotropic scenario (i.e., dynamical), with an odds ratio in favor of alignment of \aliOddsVal.  Similarly to \cite{farrnature}, but with a more general model, we find that if all of LIGO's BBHs are assumed to come from this aligned population, then most BH spins must be low ($a\lesssim 1/3$).

Figure~\ref{fig:3bin_odds_ratio} shows that $\sim10$ additional detections will be sufficient to distinguish between a pure aligned or isotropic population, unless the intrinsic spin magnitude distribution is very low (and GW151226 turns out to be an outlier in spin). LIGO is on the cusp of providing important constraints on the formation mechanisms of its binary black holes.

\begin{acknowledgements}
We thank Jonathan Gair for valuable discussions.  BF and DEH were partially supported by NSF CAREER grant PHY-1151836 and NSF grant PHYS-1708081. They were also supported by the Kavli Institute for Cosmological Physics at the University of Chicago through NSF grant PHY-1125897 and an endowment from the Kavli Foundation. The authors thank the Niels Bohr Institute for its hospitality while part of this work was completed, and acknowledge the Kavli Foundation and the DNRF for supporting the 2017 Kavli Summer Program.
\end{acknowledgements}

\bibliographystyle{apsrev4-1}
\bibliography{bibliography}
\end{document}